\documentclass[aps,showpacs,11pt,prd,tightenlines,a4,nofootinbib]{revtex4}

\usepackage{graphicx}
\usepackage{bm}
\usepackage{amsmath,amssymb}
\usepackage{latexsym}

\usepackage{upgreek}
\allowdisplaybreaks

\usepackage{color}

\begin{document}

\title{Stress-energy tensor of the quantized massive fields in  
Schwarzschild-Tangherlini  spacetimes. The back reaction.}

\author{Jerzy Matyjasek}
\email{jurek@kft.umcs.lublin.pl}
\affiliation{Institute of Physics, Maria Curie-Sk\l odowska University\\
pl. Marii Curie-Sk\l odowskiej 1, 20-031 Lublin, Poland}

\author{Pawe{\l} Sadurski}
\affiliation{Institute of Physics, Maria Curie-Sk\l odowska University\\
pl. Marii Curie-Sk\l odowskiej 1, 20-031 Lublin, Poland}

\date{\today}

\begin{abstract}
We construct and study the approximate stress-energy tensor of the quantized massive scalar field 
in higher dimensional Schwarzschild-Tangherlini spacetimes. The stress-energy tensor is calculated
within the framework of the Schwinger-DeWitt approach. It is shown that in $N$-dimensional spacetime
the main approximation can be obtained from the effective action constructed 
form the coincidence 
limit of the Hadamard-DeWitt coefficient $a_{k},$ where $k-1$ is the integer part of $N/2$. 
The back reaction of the quantized field upon the black hole spacetime is analyzed and
the quantum-corrected Komar mass and the Hawking temperature is calculated.
It is shown that for the  minimal and conformal coupling the increase of the  Komar mass 
of the quantum corrected  black hole leads to the decrease of its Hawking temperature. 
This is not generally true for more exotic values of the coupling parameter. 
The general formula describing the vacuum polarization, $ \langle \upphi^{2} 
\rangle,$ is constructed and  briefly examined.
\end{abstract}



\maketitle
\section{\label{intro}Introduction}

The aim of this paper is to construct and discuss the regularized stress-energy 
tensor of the quantized massive scalar field in a large mass limit in the 
spacetime of $N$-dimensional static and spherically-symmetric 
Schwarzschild-Tangherlini black hole described by 
the line element~\cite{Tangherlini}
\begin{equation}
 ds^{2} = -f^{(0)}(r) dt^{2} + \frac{1}{f^{(0)}(r)} dr^{2} + r^{2} d\Omega^{2}_{N-2}
\end{equation}
with
\begin{equation}
 f^{(0)}(r) = 1 - \left( \frac{r_{+}}{r}\right)^{N-3},
\end{equation}
where $d\Omega^{2}_{N-2}$ is a metric on a unit $(N-2)$-dimensional sphere,
and, subsequently, analyze its influence on the background geometry via the
semiclassical Einstein field equations. One can associate the mass with the 
solution simply by comparing its asymptotic behavior with the solutions
of  $N$-dimensional linearized gravity. The mass, $M,$ of the classical 
Schwarzschild-Tangherlini black hole is given by
\begin{equation}
 M = \frac{\pi^{(N-3)/2}(N-2)}{8 \Gamma\left( \frac{N-1}{2}\right)} r_{+}^{N-3},
 \label{mass1}
\end{equation}
where $r_{+}$ is the radial coordinate of the event horizon. The Hawking 
temperature calculated in the standard way is always inversely proportional to
the radius of the event horizon
\begin{equation}
  T_{H} = \frac{N-3}{4 \pi r_{+}}.
  \label{thawk}
 \end{equation}
The quantum-corrected  solution is, of course, characterized by a different 
radius of the event horizon and  Eqs.~(\ref{mass1}) and (\ref{thawk}) do not hold.

The Schwarzschild-Tangherlini black holes are classically stable with respect
to the linear perturbations, Moreover,  it can be
demonstrated that there are no static scalar perturbations that is regular 
everywhere outside the event horizon~\cite{Kodama}. The latter means that
if such  perturbations exist it would be possible to construct an asymptotically
flat vacuum black hole solutions  with nonspherical event horizons of
topology $S^{N-2}.$ The non-existence of such solutions confirms the uniqueness
of the $N$-dimensional spherically-symmetric static vacuum black holes.

The stress-energy tensor of the quantized field employed in this paper is 
constructed  within a generalized
Schwinger-DeWitt framework~\cite{Julian,Bryce1,dew75,Barvinsky,FZ3,Avramidi}. 
In this approach one assumes that for sufficiently massive quantized fields
the vacuum polarization effects can  be separated from the particle creation. 
Since  the vacuum polarization is local and for a given type of field it 
depends solely on the spacetime geometry, it is possible to construct the 
general expression describing the one-loop effective action.
The stress-energy tensor can be obtained by differentiating the 
effective action with respect to the metric and the result is a linear 
combination of the purely geometric terms constructed form curvature.
Moreover, as the particle creation is negligible in this regime,
the geometric approximation based on the Schwinger-DeWitt method is expected to 
be  quite good. Indeed, extensive numerical analyses carried out in 
Ref.~\cite{PaulA} indicate that for $N=4$  black holes, the relative 
error of the approximation is below $2\%$, provided $M m > 2.$ It is a very important 
result  as it explicitly demonstrates the usefulness of the method. The general 
criterion for applicability of the approximation is that  the length scale 
associated with the quantized field should be much smaller than the 
characteristic scale of the curvature of the spacetime.

The paper is organized as follows. In section II we construct the general
expression describing the stress-energy tensor of the quantized massive 
scalar fields in a large mass limit in $N$-dimensional spacetime. Subsequently, 
the general formulas are used in $N=4, 5, 6$ and $7$-dimensional 
Schwarzschild-Tangherlini spacetimes. The semi-classical Einstein field
equations are investigated in Sec. III, where the back reaction of the 
quantized fields upon the spacetime metric is examined.  
Section IV concludes the paper with some final remarks, putting our
results in a somewhat broader perspective. Also in that section the field 
fluctuation, $ \langle \upphi^{2} \rangle,$ is constructed and briefly examined.

Throughout the paper the natural system of unit is used. The signature of the 
metric is `` mainly positive'' $(-,+,...,+)$ and our conventions for curvature 
are $R^{a}_{\ bcd} = \partial_{c} \Gamma^{a}_{bd}  ...$ and $R^{a}_{\ bac} = 
R_{bc}.$

\section{The stress-energy tensor}

Let us start with the massive scalar field, $\upphi,$  propagating on $N$-dimensional 
spacetime, satisfying the covariant Klein-Gordon equation.  The associated 
Green function is the solution of the equation 
\begin{equation}
 \left(\Box - m^{2} - \xi R \right) G(x,x') =- \delta(x,x')\equiv - \frac{\delta(x-x')}{|g|^{1/2}},
\end{equation}
where $m$ is the mass of the field, $\xi$  is the parameter of the curvature 
coupling and $R$ is the curvature scalar. 

Now, making use of the (formal) definition of the one-loop effective action 
$W^{(1)}$ in the standard form 
\begin{equation}
 W^{(1)} = - \frac{i}{2} \ln {\rm Tr} G
\end{equation}
and the Schwinger-DeWitt representation of the Green function
\begin{equation}
G^{F}(x,x') =   \frac{i \Delta^{1/2}}{(4 \pi)^{n/2}} \int_{0}^{\infty} i ds 
\frac{1}{(is)^{n/2}} \exp\left[-i m^{2} s + \frac{i \sigma(x,x')}{2 s} \right] 
A(x,x'; is) ,
\label{grf}
\end{equation}
expressed in terms of the Hadamard-DeWitt coefficients, $a_{k}(x,x'),$
where $\Delta$ is the vanVleck-Morette determinant constructed form the word 
function  $\sigma$ (a biscalar equal to one half the square of the geodesic 
distance between $x$ and $x'$) and 
\begin{equation}
 A(x,x';is) = \sum_{k=0}^{\infty} (is)^{k} a_{k}(x,x'),
\end{equation}
one obtains
\begin{equation}
 W^{(1)} = \lim_{x' \to x} \int d^{N}x (-g)^{1/2}  
 \frac{\Delta^{1/2}}{2 (4 \pi)^{N/2}} \int_{0}^{\infty}    
 \frac{ids}{(is)^{N/2+1}} \exp\left[-i m^{2} s + \frac{i \sigma}{2 s} \right] A(x,x';is).
\end{equation}
Consequently, the effective Lagrangian density is given by 
\begin{equation}
 {\cal L} = \frac{1}{2(4 \pi)^{N/2}} \int \frac{i ds}{(i s)^{N/2+1} } 
 e^{-i m^{2} s} \sum_{k=0}^{\infty} a_{k} (is)^k,
\end{equation}
where $a_{k}$ is the coincidence limit of $a_{k}(x,x'),$
i.e., $a_{k} = \lim_{x'\to x} a_{k}(x,x').$

Let $\lfloor x \rfloor$ denote the floor function, i.e., it gives the largest integer less than or 
equal to $x.$
Since the first $\lfloor \frac{N}{2} \rfloor +1$ terms of the series (counting from the 
zeroth-term) lead to the divergent integrals,
let us substitute $A$ in (\ref{grf}) by its `regularized' counterpart
\begin{equation}
 A_{reg}(x,x';is) = \sum_{k= \lfloor \frac{N}{2} \rfloor +1}^{n'} a_{k}(x,x';is) (is)^{k}.
\end{equation}
The upper limit $n'$ reflects the fact that only a first few Hadamard-DeWitt
coefficients are known.

Assuming that $m^{2}$ has a small imaginary part ($i\varepsilon,$ $\varepsilon 
<0$) 
and integrating over  $s$  gives
\begin{equation}
 {\cal L}_{reg} = \frac{1}{2 (4 \pi)^{N/2}} \sum_{k= \lfloor \frac{N}{2} \rfloor 
+1}^{n'}
 \frac{a_{k}}{(m^{2})^{k-N/2}} \Gamma(k-\frac{N}{2}).
 \label{Lreg}
\end{equation}
The (regularized) stress-energy tensor can be calculated from the standard definition 
\begin{equation}
 T^{ab} = \frac{2}{(-g)^{1/2}} \frac{\delta}{\delta g_{ab}} W_{reg}^{(1)},
\end{equation}
where  $W_{reg}^{(1)}$ is given by
\begin{equation}
 W_{reg}^{(1)} = \int d^{N}x (-g)^{1/2} {\cal L}_{reg}.
 \label{Wreg}
\end{equation}
This result may be thought of as a generalization of the Frolov-Zel'nikov 
formula to the $N$-dimensional case.

In what follows we  restrict ourselves to the first-order approximation, i.e.,
for a given $N$ we retain only the lowest regular term of the expansion
(\ref{Lreg}) and denote resulting Lagrangian density by $ L^{N}$. 
Inspection of (\ref{Lreg}-\ref{Wreg}) shows that to calculate the\
approximate  stress-energy tensor in the spacetimes of dimension 4 and 5 the
coincidence limit of the fourth coefficient, $a_{3},$ is needed.
Similarly, the coefficient $a_{4}$ allows calculations in $N=6$  and 7,  and so
on. Unfortunately, the Hadamard-DeWitt coefficients, except for simple
geometries with a high degree of symmetry, are very  hard to calculate as they
are constructed from the differential and algebraic curvature invariants.
The differential invariants involve the covariant derivatives of the curvature
tensor (and their contractions) up to $(n-2)$-order~\cite{sakai,gilkey,Amsterdamski,Avramidi,Ven}.
The problem at hand is even
more complicated, since what we need is the result of the functional
differentiation of the (integrated) coefficient $a_{k}$  with respect to the
metric tensor rather than the coefficient itself. To make things  worse, we have
to apply the thus obtained formulas in a concrete spacetime, what is usually
associated with  large-scale calculations. 

Before going any further, let us summarize what has been done so far. 
Here we limit ourselves almost exclusively to literature on the regularized
stress-energy tensor calculated within the framework of the Schwinger-DeWitt 
approximation.  Assuming that the Compton length associated with the mass 
of the field is much less than the gravitational radius of the  black hole,
Frolov and Zel'nikov~\cite{FZ1} constructed the stress-energy tensor 
of the massive scalar field in the Hartle-Hawking state 
in the Schwarzschild spacetime.  The large mass limit allows separation of the
vacuum polarization effects and the final result can be calculated from the
(coincidence limit) of the Ricci-flat version of the coefficient $a_{3}.$ The
scalar results have been extended to spin $1/2$ and spin $1$ fields in the Kerr
spacetime~\cite{FZ2,FZ3}. The Forolv-Zel'nikov results (for all mentioned spins)
have been generalized to {\it arbitrary} spacetime in Refs.~\cite{ja1,ja2}.
This has been achieved by constructing the functional derivatives of 10
curvature (algebraic and differential) invariants of the background 
dimensionality 6
(i.e. having the dimension of $length^{-6}$) with respect 
to the metric tensor. 
In the $N=4$  case, the resulting
stress-energy tensor consists of almost 100 geometric terms 
constructed from the curvature and metric. Interested reader
in referred to Refs.~\cite{ja1,ja2}. 
Identical results for the static spherically-symmetric 
asymptotically-flat geometries have been obtained using different methods in 
Ref~\cite{Samuel}.  The analysis of the functional derivatives of the curvature 
invariants have been also carried out by Decanini and Folacci in Refs~\cite{Dec1,Dec2}.  
A natural question that appears in this context is the problem of the quality
of the approximation.  A detailed numerical study carried out in Ref.~\cite{PaulA} indicates
that the Schwinger-DeWitt approximation, when employed in its domain of applicability,
is reliable.  

The stress-energy tensor have been calculated in numerous, physically interesting 
geometries, such as exterior and interior regions of black holes~\cite{shane,jmP}, 
wormholes~\cite{Taylor} and cosmology~\cite{frwl2013,frwl2014}. Interesting results 
have been obtained in the geometries with maximally symmetric subspaces, such as 
the Bertotti-Robinson solution~\cite{Kofman1,Kofman2,Mata42}. Recently, there 
is a growing interest in the higher dimensional calculations, 
(see e.g.,~\cite{LemosT,Kent} and the references cited therein), that reflects
the view that the physical world has more than the familiar four dimensions.

Now, let us return to our main problem. To construct the first-order
approximation to the stress-energy tensor one has to calculate the variational
derivatives of the effective action expressed in terms of the coincidence limit
of the heat kernel coefficients for arbitrary dimensions. Here we shall limit
ourselves to coefficients $a_{3}$ and $a_{4}.$  Using FORM, which is particularly suited 
for large
scale calculations~\cite{Vermaseren,Tentyukov,Kuipers},  we have constructed 
the coincidence limit of the
coefficients $a_{3}$ and $a_{4}$ and subsequently the functional derivatives of
the effective action with respect to the metric tensor.  After some
simplifications we have obtained the general expressions (stored in FORM format)
describing the stress-energy tensor of the quantized massive scalar field in
$N=4,5,6$ and 7-dimensional geometries, respectively. Unfortunately, the general
results are very complicated, and, except for the geometries with a high degree
of symmetry, hard to use. 

In the light of the foregoing discussion, to shorten the presentation 
and minimize efforts, here we will 
follow a less general approach~\footnote{It should be noted however, that all calculations of
the stress-energy tensor presented in this paper have been checked using this
more general approach.}. The static spherically symmetric solution 
of the Einstein field equations, written in the  standard curvature
coordinates, has the form
\begin{equation}
ds^{2} = g_{00}(r) dt^{2} + g_{11}(r) dr^{2} + r^{2} d \Omega^{2}_{N-2},
\label{elx}
\end{equation}
where $d \Omega^{2}_{N-2}$ is the line element on a unit sphere $S^{N-2}.$ To 
simplify notation, let us introduce two functions 
$f(r) $ and $h(r)$ defined as $f(r) =g_{00}(r),$ and $h(r)= g_{11}(r),$ 
respectively. Calculating the Hadamard-DeWitt 
coefficient for the line element one obtains the Lagrangian density, $L^{N},$
which can be schematically written 
in the form 
\begin{equation}
L^{N} = {\cal L}^{N}\left(f(r),...,f^{(i_{N})}(r),h(r),...,h^{(j_{N})}(r),r\right) \sqrt{g_{S_{N-2}}},
                                          \label{Lagr_c}
\end{equation}
where and $g_{S_{N-2}}$ is the determinant of the metric tensor on a unit $S_{N-2}$ sphere,
 $f^{(k)}$  and $h^{(k)}$ denote a $k-$th derivative of $f(r)$ and
$h(r),$ respectively. Note that the  numerical coefficient, the mass and the 
factor $\sqrt{f(r) h(r)}$
 have been absorbed into the definition of ${\cal L}^{N}.$
Now the stress-energy tensor can be obtained from the Euler-Lagrange equations
\begin{equation}
T_{t}^{(N)t} =2 \left(\frac{f}{h}\right)^{1/2}
\left[ \frac{\partial}{\partial f}{\cal L}^{N} 
+  \sum_{k=1}^{p(N)}\left(-1\right)^{k} 
\frac{d^{k}}{dr^{k}} \left(\frac{\partial}{\partial f^{(k)}} {\cal L}^{N}\right) \right]
      \label{t_comp}
\end{equation}
and
\begin{equation}
T_{r}^{(N)r} =2 \left(\frac{h}{f}\right)^{1/2}
\left[\frac{\partial}{\partial h}{\cal L}^{N} 
+ \sum_{k=1}^{s(N)}\left(-1\right)^{k} \frac{d^{k}}{dr^{k}} 
\left(\frac{\partial}{\partial h^{(k)}} {\cal L}^{N}\right) \right],
      \label{r_comp}
\end{equation}
where $p(N)$ and $s(N)$ can easily be inferred form the Lagrangian density.
The angular components can be obtained from the covariant conservation equation 
$\nabla_{a} T^{ab} = 0,$ which, for the line element (\ref{elx}), reduces to
\begin{equation}
T^{(N)\alpha_{1}}_{\alpha_{1}}=... = T^{(N)\alpha_{N-2}}_{\alpha_{N-2}} = 
-\frac{r}{2 f(N-2)} 
\left(T^{(N)t}_{t} -T^{(N)r}_{r} \right)\frac{d }{dr} f +
\frac{r}{N-2}\frac{d }{dr}T^{(N)r}_{r}  + T^{(N)r}_{r},
\label{katowa}
\end{equation} 
where $T^{(N)\alpha_{i}}_{\alpha_{i}}$ is any angular component of the 
stress-energy tensor. The coordinates $\{\alpha_{1},...,\alpha_{N-2}\}$ cover
the $N-2$-dimensional sphere. Note that once the time and radial components of the stress-energy tensor are known
the angular components can be obtained at practically no expense.

Making use of the coincidence limit of the Hadamard-DeWitt coefficient 
$a_{3}(x,x')$ in the  $N=4$ case, one has
\begin{equation}
 T^{(4)t}_{t} = \frac{1}{ m^2 \pi^2 r^6 }\left[\frac{1237 x^3}{40320}-\frac{25 
x^2}{896}+\left(\frac{x^2}{8}-\frac{11 x^3}{80}\right) \xi\right] 
 \label{firstcomp}
\end{equation}
and
\begin{equation}
 T^{(4)r}_{r} = \frac{1}{ m^2 \pi^2 r^6 }\left[ -\frac{47 x^3}{5760}+\frac{7 
x^2}{640}+\left(\frac{3 x^3}{80}-\frac{x^2}{20}\right) \xi\right], 
\end{equation}
where $x= r_{+}/r$.
Similarly, for $N=5$ one obtains
\begin{equation}
 T^{(5)t}_{t} = \frac{1}{ m \pi^2 r^6 }\left[ \frac{841 x^6}{5040}-\frac{81 
x^4}{560}+\left(\frac{3 x^4}{5}-\frac{7 x^6}{10}\right) \xi\right]
\end{equation}
and
\begin{equation}
T^{(5)r}_{r} = \frac{1}{ m \pi^2 r^6 }\left[ -\frac{37 x^6}{1008}+\frac{33 
x^4}{560}+\left(\frac{x^6}{5}-\frac{3 x^4}{10}\right)\xi\right].
\end{equation}
The calculations of the stress-energy tensor in the spacetime of the higher-dimensional black holes require 
the knowledge of the higher-order  Hadamard-DeWitt coefficients. Indeed, making use of the coincidence limit of 
the coefficient $a_{4}(x,x')$ in  $6$-dimensional Schwarzschild-Tangherlini spacetime  gives
\begin{equation}
 T^{(6)t}_{t}  =
\frac{1}{m^{2} \pi^{3} r^{8}} 
\left[ -\frac{73973 x^{12}}{5040}+\frac{40457 x^9}{2016}-\frac{387 x^6}{64}
+\xi  \left(\frac{59985 x^{12}}{896}-\frac{19945 x^9}{224}+\frac{405 
x^6}{16}\right) \right]
\end{equation}
and
\begin{equation}
  T^{(6)r}_{r} =  \frac{1}{m^{2} \pi^{3} r^{8}} \left[ \frac{26969 
x^{12}}{10080}
  -\frac{103 x^9}{18}+\frac{153 x^6}{64}+\xi  \left(-\frac{33325 
x^{12}}{2688}+\frac{18055 x^9}{672}-\frac{45 x^6}{4}\right)\right],
\end{equation}
whereas for $N=7$ one obtains
\begin{equation}
  T^{(7)t}_{t} =   \frac{1}{m \pi^{3} r^{8}}  \left[-\frac{4713  x^{16}}{128}
  +\frac{387  x^{12}}{8}-\frac{217  x^8}{16}+\xi  \left(\frac{1188  
x^{16}}{7}-216  x^{12}+\frac{225  x^8}{4}\right)\right]
\end{equation}
and
\begin{equation}
 T^{(7)r}_{r} =   \frac{1}{m \pi^{3} r^{8}}  \left[ \frac{30549  x^{16}}{4480}
 -\frac{8261  x^{12}}{560}+\frac{237  x^8}{40}+\xi  \left(-\frac{891  
x^{16}}{28}+\frac{3915  x^{12}}{56}-\frac{225  x^8}{8}\right) \right].
\label{lastcomp}
 \end{equation}
 The components of the stress-energy tensor in 4-dimensional Schwarzschild spacetime
 have been calculated earlier in Refs.~\cite{FZ1,FZ2}.
As the angular components of the stress-energy tensor can easily be calculated 
form  the covariant conservation equation~(\ref{katowa}) we shall not display 
them here. 

The intermediate calculations of $ T^{(N)a}_{b}$  are rather 
complicated but the final result is surprisingly simple, with only a weak 
increase of its complexity with dimension. It should be noted that in general, 
the coincidence limit of $a_{k}$ is a $k-$th degree polynomial in $\xi,$ with 
the (geometric) coefficients of $\xi^{i},$ for $i >1,$ involving products of the 
Ricci tensor, its contractions and covariant derivatives. Additionally,
there is a term $\Box^{k-1}R,$ which, being a total divergence, does not 
contribute to the final result. That explains why the stress-energy tensor in 
the Schwarzschild-Tangherlini spacetime is always linear in $\xi.$ The same is 
true for the more general Ricci-flat metrics. This behavior can be easily 
traced back to the recurrence equation for the general Hadamard-DeWitt 
coefficient $a_{k}(x,x').$

The stress-energy tensor is regular in a physical sense if it is regular in  
a freely-falling frame of reference. To demonstrate that the components of the 
stress-energy tensor (\ref{firstcomp}-\ref{lastcomp}) do satisfy this 
requirement let us introduce the vectors of the frame defined as follows. For 
radial motion the frame consists of the $N$-velocity vector 
$e^{a}_{(0)} =u^{a}$ 
and a unit length spacelike vector $e^{a}_{(1)} = n^{a}.$  (The remaining 
vectors of the frame are unimportant for our purposes). Now, using the geodesic 
equations, one has
\begin{equation}
 e^{a}_{(0)} = u^{a} = \left(\frac{E_{0}}{f} , \-\sqrt{\left( 
\frac{E_{0}^{2}}{f}-1 \right)\frac{1}{h}},0,...,0\right)
 \label{fistcomp}
\end{equation}
and
\begin{equation}
 e^{a}_{(1)} = n^{a} = \left( -\frac{1}{f} \sqrt{E_{0}^{2}-f   }, \frac{E_{0}}{\sqrt{fh}},0,...,0\right),
\end{equation}
where $E_{0} $ is the constant of motion. The components of the stress-energy tensor 
in the frame can be written in the form:
\begin{equation}
 T_{(0)(0)} = -\frac{ E_{0}^{2}\left(T^{0}_{0}-T^{1}_{1} \right)}{f} - T^{1}_{1}
\end{equation}
\begin{equation}
 T_{(1)(1)} = -\frac{ E_{0}^{2}\left(T^{0}_{0}-T^{1}_{1} \right)}{f} + T^{1}_{1}
\end{equation}
\begin{equation}
 T_{(0)(1)} = \frac{ E_{0}\sqrt{E_{0}^2-f}\left(T^{0}_{0}-T^{1}_{1} \right)}{f} ,
\end{equation}
and, consequently, the stress-energy tensor in a freely-falling frame is regular as $r\to r_{+}$
if 
\begin{equation}
|T_{a}^{b}| <\infty \hspace{0.5cm} {\rm  and} \hspace{0.5cm} |\left(T^{0}_{0}-T^{1}_{1} \right)/f|<\infty.
\end{equation}
Inspection of (\ref{firstcomp}-\ref{lastcomp}) shows that 
\begin{equation}
 T^{(N)r}_{r} -T^{(N)t}_{t} = F(r) \left[1 -\left(\frac{r_{+}}{r}\right)^{N-3}\right],
\label{factorization}
 \end{equation}
where $F(r)$ is a simple polynomial in $r_{+}/r,$ 
and, consequently, the components $T_{(0)(0)}$ $T_{(1)(1)}$ and $T_{(0)(1)}$
are regular. Moreover, by the same argument, the components  
$T^{(N)\alpha_{i}}_{\alpha_{i}}$ given by Eq. (\ref{katowa}) are regular also. 
We would like to emphasize that as the tensors have been  calculated using 
various  computational strategies, the regularity of the angular components  
has been established independently.

Although interesting in its own right, the main role played by the 
stress-energy tensor is to provide the source term to the semiclassical 
Einstein field equation. The back reaction of the quantized fields upon the 
classical background is the main theme of the next section.
\section{The Back reaction}

In their simplest form the semiclassical Einstein field equations can be written 
as
\begin{equation}
 G^{a}_{b} = 8\pi T^{(N)a}_{b},
\end{equation}
where, in general, the total stress energy tensor describes both classical and 
quantum matter. Ideally, the stress-energy tensor of the quantized field should 
functionally depend on a general metric or at least on the wide class of 
metrics. This allows, in principle, to construct the solution of the 
semiclassical Einstein field equations in a self-consistent way. On the other 
hand, one can follow a simpler approach, in which the stress-energy tensor 
is calculated in a concrete spacetime and the back reaction on the metric
is treated perturbatively. In the black hole context the semiclassical
Einstein field equations have been studied for the first time by 
York~\cite{York} more than thirty years ago (see also Ref.~\cite{Sanchez}).
Since then various aspects of the back reaction problem have been studied in a 
number of papers, see 
e.g.,~\cite{Hochberg1,Hochberg2,Fursaev,Banerjee,Banerjee2,Tryniecki,Kasia} and 
the references cited therein.

In order to construct the semi-classical Einstein field equations, let us 
start with the line element 
\begin{equation}
 ds^{2} = -f(r) dt^{2} + h(r) dr^{2} + r^{2} d\Omega^{2}_{N-2} ,
 \label{ddgen}
\end{equation}
where
\begin{equation}
 f(r) = e^{2 \psi(r)} \left(  1 - \frac{2 M(r)}{r^{N-3}}\right)  \hspace{.5cm} 
{\rm and} \hspace{.5cm} h(r) =  \left(  1 - \frac{2 M(r)}{r^{N-3}}\right)^{-1}.
\end{equation}
The main reason for introducing the new functions $M(r)$ and $\psi(r)$ is to 
simplify the resulting equations. 
With such a substitution, the semiclassical Einstein field equations read 
\begin{equation}
 \frac{dM}{dr} = - \varepsilon \frac{8\pi r^{N-2}}{N-2} T^{(N)t}_{t}
 \label{diffM}
\end{equation}
and
\begin{equation}
 \frac{d\psi}{dr} = \varepsilon \frac{8 \pi r}{N-2} \frac{T^{(N)r}_{r} - 
T^{(N)t}_{t}}{1 -\frac{2 M}{r^{N-3}}},
 \label{diffPsi}
\end{equation}
where to simplify the calculations and to keep control of the order of terms in the complicated
series expansions we have introduced the dimensionless parameter, $\varepsilon,$ substituting
$T^{(N)a}_{b} \to \varepsilon T^{(N)a}_{b}.$ We have to put $\varepsilon=1$ at the final stage of calculations.

The quantum corrections to the Schwarzschild-Tangherlini metric can be 
calculated making use of the  expansion  
\begin{equation}
 M(r) = \frac{r_{+}^{N-3}}{2}\left[1 + \varepsilon (N-3) \mu(r)  \right]
\end{equation}
in (\ref{diffM}) and (\ref{diffPsi}), and integrating the linearized equation 
with the initial condition $ \mu(r_{+}) =C_{1}.$ This condition means 
that the function $\mu(r)$ can be written as $\mu(r) = \mu_{0}(r) + C_{1}$ with 
$\mu_{0}(r_{+}) =0.$ The second equation can easily be integrated with the 
natural condition $\psi(\infty) = 0.$ Note that with such a choice  $\psi(r) 
\sim {\cal O}(\varepsilon).$ Putting this all together one has
\begin{equation}
 f(r) = 1 - \left( \frac{r_{+}}{r}\right)^{N-3} \left(1 + \varepsilon (N-3) 
C_{1} \right) 
 - \varepsilon (N-3) \left( \frac{r_{+}}{r}\right)^{N-3} \mu_{0}(r),
 \label{fodr}
\end{equation}
where 
\begin{equation}
 \mu_{0}(r) = -\frac{16 \pi}{ r_{+}^{N-3} (N-3)(N-2)} 
 \int_{r_{+}}^{r} r^{N-2} T^{(N)t}_{t} dr .
\end{equation}
The integration constant $C_{1}$ can be absorbed into the definition of the  radius of the event 
horizon $r_{H}$ as follows 
\begin{equation}
  r_{H} = r_{+}(1 + \varepsilon C_{1})
\end{equation}
in the process of  the finite renormalization. The physical radius of the event 
horizon, $r_{H},$ is measurable as opposed to the unphysical (bare) $r_{+}.$   
Since $\mu_{0}$ depends on $r$ and $r_{+}$ and   the third
term on the right hand side of Eq. (\ref{fodr}) is ${\cal O }(\varepsilon),$ in 
the linearized calculations, 
one can use $r_{H}$ instead of $r_{+}$ both in $\mu(r)$ and $\psi(r).$  
With such a substitution one introduces ${\mathcal{O}}(\varepsilon^{2})$ error. 
Let us return to the second equation of the system. Since 
Eq.~(\ref{factorization}) holds,  
the problem reduces to the two simple quadratures.

The same result can be obtained solving the semiclassical  Einstein field  
equations  
with the stress-energy tensor depending on a general metric and with the 
quantum-corrected `exact' event horizon, $r_{H},$ 
as the initial condition from the very beginning.~\footnote{To 
cross-check the calculations we have employed both methods.}  
Let us employ  the second method and construct the semiclassical Einstein 
field equations for $M(r)$ and $\psi(r)$ with the initial conditions 
\begin{equation}
 M(r_{H}) = \frac{1}{2} r_{H}^{N-3} \hspace{0.5cm} {\rm and} \hspace{0.5cm} \psi(\infty) =0.
\end{equation}
Assuming 
\begin{equation}
 M(r) = M_{0}(r) + \varepsilon M_{1}(r) + {\cal O}(\varepsilon^{2}) 
\hspace{0.5cm}{\rm and} \hspace{0.5cm} \psi(r) = \varepsilon \psi_{1}(r) +  
{\cal O}(\varepsilon^{2})
 \label{emf}
\end{equation}
one obtains differential equations which can be solved with the conditions
\begin{equation}
 M_{0}(r_{H}) =  \frac{1}{2} r_{H}^{N-3}, \hspace{0.5cm} M_{1}(r_{H}) =  0 \hspace{0.5cm}{\rm and} \hspace{0.5cm} \psi_{1}(\infty)= 0.
\end{equation}
The zeroth-order equation for a general $N$ gives
\begin{equation}
 M_{0}(r) = \frac{1}{2} r_{H}^{N-3},
\end{equation}
whereas the functions $M_{1}(r)$ and $\psi_{1}(r)$ assume more complicated, dimension-dependent form.
After some algebra, one has  
\begin{equation}
 M_{1}(r) = \frac{1}{\pi m^{2}} \left[\frac{1237 r_{H}^3}{60480 r^6}-\frac{5 r_{H}^2}{224 r^5} +\frac{113}{60480 r_{H}^3} +\xi  \left(-\frac{11 r_{H}^3}{120 r^6}
 +\frac{r_{H}^2}{10 r^5}-\frac{1}{120 r_{H}^3}\right)\right],
\end{equation}
\begin{equation}
 \psi_{1}(r) = \frac{1}{\pi m^{2}}\left(  \frac{7 r_{H}^2 \xi }{60 r^6}-\frac{29 r_{H}^2}{1120 r^6}\right)
\end{equation}
and
\begin{equation}
 M_{1}(r) = \frac{1}{\pi m}\left[\frac{841 r_{H}^6}{15120 r^8}-\frac{9 
r_{H}^4}{140 r^6}+ \frac{131}{15120 r_{H}^2} 
 + \xi  \left(-\frac{7 r_{H}^6}{30 r^8}+\frac{4 r_{H}^4}{15 r^6}-\frac{1}{30 
r_{H}^2}\right)\right],
\end{equation}
\begin{equation}
 \psi_{1}(r) = - \frac{1}{\pi m} \left( \frac{19 r_{H}^4}{280  r^8}  -  \frac{3 r_{H}^4  }{10  r^8} \xi  \right),
\end{equation}
respectively for $N=4$ and $N=5.$
The analogous calculations in higher dimensional spacetimes are slightly more 
involved and for $N=6$ give
\begin{eqnarray}
M_{1}(r) &=& \frac{1}{\pi^{2} m^{2}} \left[ -\frac{73973 r_{H}^{12}}{37800 
r^{15}}
+\frac{40457 r_{H}^9}{12096 r^{12}}-\frac{43 r_{H}^6}{32 r^9}  
-\frac{13291}{302400 r_{H}^3} \right.
 \nonumber \\
&& \left. +  \xi  \left(\frac{3999 r_{H}^{12}}{448
   r^{15}}-\frac{19945 r_{H}^9}{1344 r^{12}}+\frac{45 r_{H}^6}{8 r^9}+\frac{97}{336 r_{H}^3}\right)\right]
\end{eqnarray}
and
\begin{equation}
 \psi_{1}(r) = \frac{1}{\pi^{2} m^{2}}\left[\frac{3887 r_{H}^9}{1680 
r^{15}}-\frac{45 r_{H}^6}{32 r^{12}}
 +\xi  \left(\frac{195 r_{H}^6}{32 r^{12}}-\frac{1333 r_{H}^9}{126 
r^{15}}\right)\right].
\end{equation}
Similarly, for $N=7$ one has
\begin{eqnarray}
M_{1}(r) &=& \frac{1}{\pi^{2} m}
\left[-\frac{1571 r_{H}^{16}}{480 r^{18}}+\frac{387 r_{H}^{12}}{70 r^{14}}
-\frac{217 r_{H}^8}{100 r^{10}} -\frac{1439}{16800 r_{H}^{2}}\right. \nonumber 
\\
&&\left. +\xi  \left(\frac{528 r_{H}^{16}}{35 r^{18}}-\frac{864
   r_{H}^{12}}{35 r^{14}}+\frac{9 r_{H}^8}{r^{10}}+\frac{3}{5 r_{H}^2}\right) \right]
\end{eqnarray}
and
\begin{equation}    
 \psi_{1}(r) = \frac{1}{\pi^{2} m}\left[\frac{4073 r_{H}^{12}}{1050 
r^{18}}-\frac{1559 r_{H}^8}{700 r^{14}}+\xi  \left(\frac{135 r_{H}^8}{14 
r^{14}}-\frac{627 r_{H}^{12}}{35 r^{18}}\right)\right].
\end{equation}

Having established the form of the quantum corrected metric the correction to 
the temperature of the Schwarzschild-Tangherlini black hole  can be calculated. 
First, observe that for the static and spherically symmetric black hole the 
Euclidean version of the line element has no conical singularity, provided  
the complexified time coordinate is periodic with a period $\beta$ given by
\begin{equation}
 \beta =\lim_{r\to r_{H}} 4\pi  (g_{00} g_{11})^{1/2} \left( \frac{d}{dr} g_{00} \right)^{-1}.
\end{equation}
Thus, as in the classical Schwarzschild-Tangherlini spacetime, a quantum-corrected black hole have a natural temperature 
associated with it.
The Hawking temperature is given by $T_{H} =\beta^{-1}$ and to ${\mathcal O}(\varepsilon),$ one has 
\begin{equation}
 T^{(N)}_{H} = \frac{N-3}{4\pi r_{H}} + \varepsilon \Delta T^{(N)}_{H},
\end{equation}
where
\begin{equation}
 \Delta T^{(4)}_{H} = \frac{1}{\pi^{2} m^2 r_{H}^{5}} \left( \frac{\xi }{240}-\frac{37}{40320} \right),
\end{equation}
\begin{equation}
\Delta T^{(5)}_{H} = \frac{1}{\pi^{2} m r_{H}^{5}} \left( \frac{\xi }{60}-\frac{13}{3024}\right),
\end{equation}
\begin{equation}
 \Delta T^{(6)}_{H} = \frac{1}{\pi^{3} m^{2} r_{H}^{7}} \left( \frac{47}{1920}-\frac{97 \xi }{672} \right),
\end{equation}
\begin{equation}
 \Delta T^{(7)}_{H} = \frac{1}{\pi^{3} m r_{H}^{7}} \left( \frac{767 }{16800} -\frac{3}{10} \xi  \right).
\end{equation}
The corrections $\Delta T_{H}^{(N)}$ are linear functions of $\xi$ and one expects that this
behavior persists also in the back reaction on a more general (classical) Ricci-flat black hole geometries. 

Now, let us analyze the mass of the black hole as seen by a distant observer. 
It is evident that the mass as given by Eq.~(\ref{mass1}) is not the mass that
would be measured at great distances from the corrected black hole.
The coordinate independent 
Komar mass, $M_{\infty},$   defined by~\cite{Myers}
\begin{equation}
 \oint_{\infty} \nabla^{a} K_{(t)}^{b} d\sigma_{ab} =-16 \pi \frac{N-3}{N-2} M_{\infty}  ,
\end{equation}
where $K_{(t)}$ is the timelike Killing vector and the integrals are to be 
calculated over $(N-2)$-sphere at spatial infinity, is very useful in this regard.
Here, the Komar mass is the total mass energy of the black hole and the vacuum 
polarization of the quantized massive field. 
Making use of this definition, one has
\begin{equation}
 M_{\infty} = \frac{\pi^{(N-3)/2}(N-2)}{8 \Gamma\left( \frac{N-1}{2}\right)} r_{H}^{N-3} + \Delta M^{(N)},
\end{equation}
where
\begin{equation}
  \Delta M^{(4)} = \frac{1}{ \pi  m^2 r_{H}^3}\left(\frac{113}{60480 }-\frac{\xi }{120 } \right),
  \label{disc}
\end{equation}
\begin{equation}
  \Delta M^{(5)} = \frac{1}{m r_{H}^2}\left(\frac{131}{20160 }-\frac{\xi }{40  } 
\right),
\end{equation}
\begin{equation}
 \Delta M^{(6)} = \frac{1}{\pi  m^2 r_{H}^3}\left(\frac{97 \xi }{252 }-\frac{13291}{226800 }\right),
\end{equation}
and
\begin{equation}
  \Delta M^{(7)} =\frac{1}{m r_{H}^2}\left( \frac{3 \xi }{8 }-\frac{1439}{26880 } \right).
  \label{ddi}
\end{equation}
Precisely the same result can be easily calculated form 
\begin{equation}
 M_{\infty} = \frac{\pi^{(N-3)/2}(N-2)}{4 \Gamma\left( \frac{N-1}{2}\right)} 
\lim_{r\to \infty} M(r),
\end{equation}
where $M(r)$ is given by (\ref{emf}). 

It should be noted, however, that for $N=4,$ Eq.~(\ref{disc}) does not coincide 
with the  
result obtained by Frolov and Zel'nikov in Ref.~\cite{FZ2}, although the Komar 
mass $M_{\infty}$ is identical. 
It is simply because they used the equivalent representation for the Komar mass
\begin{equation}
 -16 \pi \frac{N-3}{N-2} M_{\infty}   = 2 \int_{S} R^{a}_{b}\, K_{(t)}^{b} d 
S_{a} + \oint_{\cal{H}} \nabla^{a} K_{(t)}^{b} \,d\sigma_{ab}, 
\end{equation} 
where ${\cal H}$ is a spatial $(N-2)$-sphere on the event horizon and $S$ is the 
region between ${\cal H}$ and space-like infinity, 
and interpreted (in $4$-dimensional spacetime) the first term on the right hand 
side of the above equation as 
 $- 8 \pi\Delta M^{(4)}_{BH}.$ Indeed, simple calculations reproduce the 
Frolov-Zel'nikov result 
 \begin{equation}
  \Delta M^{(4)}_{BH} = \frac{1}{540 \pi m^{3} r_{H}^{3}}(2-9 \xi).
 \end{equation}
On the other hand, the last term
\begin{equation}
  M_{H}= -\frac{N-2}{16 \pi (N-3)} \oint_{\cal{H}} \nabla^{a} K_{(t)}^{b} 
\,d\sigma_{ab}
\end{equation}
interpreted as a horizon-defined  black hole mass,
 when restricted to $N=4,$ gives
\begin{equation}
 M_{H} = \frac{r_{H}}{2} + \frac{1}{\pi m^{2} r_{H}^{3}} \left(\frac{\xi}{120} 
-\frac{36}{21160} \right). 
\end{equation}
It can easily be shown that the sum $\Delta M^{(4)}_{BH} +  M_{H} $ is precisely 
the Komar mass, $M_{\infty},$ of the 4-dimensional quantum-corrected 
Schwarzschild black hole.  
Both definitions of the mass correction terms have their merits and the calculation
of $\Delta M_{BH}$ presents no problem,
but, in our opinion, Eqs.~(\ref{disc}-\ref{ddi}) are better suited for further analysis.

Now, we shall analyze the influence of the quantized field on the black hole. To 
this end let us compare  
the classical and the quantum corrected black  holes, both  characterized by the 
{\it same} radius of the event horizon, $r_{H}.$ 
Two particular values of $\xi$ are of special interest:  $\xi =0,$ which 
characterizes the minimal coupling and $\xi =(N-2)(4N-4)$ 
which characterizes the conformal coupling. Other values of the coupling 
parameter are of somewhat lesser interest. 
The corrections of the Hawking temperature caused by the quantum field depend on 
the dimension and the coupling parameter and are tabulated in Table~\ref{re1}.   
  
\begin{center}
\begin{table} [h]
\begin{tabular}{l*{4}{c}}
\hline\hline
 & $N=4$ & $N=5$ & $ N=6$ & $N=7$  \\
 \hline
 $\xi =0$ &       $-$ & $-$ & $+$ & $+$ \\
 $\xi=\xi_{c} $ & $-$ & $-$ & $-$ & $-$\\
 \hline\hline
\end{tabular}
\caption{The sign of $\Delta T^{(N)}$ for two physical choices of the coupling 
parameter  
$\xi=0$ (minimal coupling) and $\xi_{c}=(N-2)/(4 N-4)$ (conformal coupling). }
\label{re1}	
\end{table}
\end{center}
Within the adopted approximation, the conformally coupled massive fields tend to 
lower the black hole temperature. On the other 
hand, under the influence of  the minimally coupled fields the Hawking 
temperature increases for  $N=4$ and $N=5$ and decreases for $N=6$ and $N=7.$ 
Similarly, inspection of Table~\ref{re2} shows that the correction to the black 
hole mass is always positive for the conformally coupled fields, whereas   
it is negative for the minimally coupled field in $N=6$ and $N=7$ dimensional 
quantum-corrected Schwarzschild-Tangherlini spacetime.  
Qualitatively, one has the following behavior for both values of the curvature coupling:
Increase of the mass of the black hole due to quantum effects decreases the 
Hawking temperature.  
It should be noted however, that for more exotic values of the parameter $\xi$ 
this observation may not necessarily be true. Finally, observe that the 
modifications of the characteristics of the black hole is bigger for minimally 
coupled fields, as can be easily seen in Table~\ref{re3}. Once again, we 
observe that for other values of the coupling parameter corrections to the 
mass and the temperature can be quite significant.

For $s$ fields with masses $m_{i}$ the main approximation to the one-loop 
effective action is still of the form (\ref{Wreg}) with $n' = \lfloor 
N/2\rfloor+1,$ 
provided the following substitution is made
\begin{equation}
 \frac{1}{(m^{2})^{\lfloor N/2\rfloor -N/2+1} }\to \sum_{i}^{s}  
\frac{1}{(m_{i}^{2})^{\lfloor N/2\rfloor -N/2+1}}. 
\end{equation}
Thus the quantum effects can be made arbitrary large  by taking a large number 
of massive fields.

\begin{center}
\begin{table} [h]
\begin{tabular}{l*{4}{c}}
\hline\hline
 & $N=4$ & $N=5$ & $ N=6$ & $N=7$  \\
 \hline
 $\xi =0$       & $+$ & $+$ & $-$ & $-$ \\
 $\xi=\xi_{c} $ & $+$ & $+$ & $+$ & $+$\\
 \hline\hline
\end{tabular}
\caption{The sign of $\Delta M^{(N)}$ for two physical choices of the coupling parameter 
$\xi=0$ (minimal coupling) and $\xi_{c}=(N-2)/(4 N-4)$ (conformal coupling). }
\label{re2}	
\end{table}
\end{center}

\begin{center}
\begin{table} [h]
\begin{tabular}{c*{4}{c}}
\hline\hline
 & $N=4$ & $N=5$ & $ N=6$ & $N=7$  \\
 \hline
 $|\Delta T_{0}/\Delta T_{c}|$       & $4.11$ & $3.66$ & $5.58$ & $2.7$ \\
 $|\Delta M_{0}/\Delta M_{c}|$         & $3.9$ & $3.59$ & $3.19$ & $2.18$\\
 \hline\hline
\end{tabular}
\caption{The (absolute) value of the ratio of $\Delta T_{0}$ to $\Delta T_{c}$ 
(the first row) and $\Delta M_{0}$ to $\Delta M_{c}$ (the second row) 
for the quantum corrected Schwarzschild-Tangherlini black hole. The minimally 
coupled field leads to more prominent corrections.} 
\label{re3}	
\end{table}
\end{center}

\section{Final remarks}

We have constructed the approximate stress-energy tensor of the quantized 
massive scalar fields in the spacetimes of the Schwarzschild-Tangherlini black 
holes. The general expressions describing the stress-energy tensor constructed 
form the coefficient $a_{3}$ ($N=4$ and $N=5$) and from $a_{4}$ ($N=6$ and 
$N=7$) have been calculated using FORM. The coefficients $a_{k}$ have 
been calculated within the framework of the manifestly covariant method.
Unfortunately, the final results
(which are  valid in any spacetime provided the applicability 
conditions are satisfied)
are rather complicated and their practical use may be limited to simple 
geometries of high symmetry. 
Although the Schwarzschild-Tangherlini black holes belong to the class of 
geometries for which such calculations can be performed 
in a reasonable time, here, for brevity, we followed a simplified approach and 
calculated the functional derivatives of the one-loop effective action
with respect to the metric potentials of the general static and spherically 
symmetric metric.
 
 Our general formulas have already been successfully tested.
 Indeed, recently we have calculated the stress-energy tensor of the quantized 
massive field in $N$-dimensional spatially-flat Friedman-Robertson-Walker 
spacetimes within the framework of the adiabatic approximation 
and it has been explicitly demonstrated that it coincides with the tensors 
obtained form the Schwinger-DeWitt method.

Finally observe, that as a by-product of the present calculations
one can easily construct the field fluctuation. Indeed, from the formal 
definition
\begin{equation}
 \langle \upphi^{2} \rangle_{reg} = - i \lim_{x' \to x}  G^{(N)}_{reg},
\end{equation}
where $ G^{(N)}_{reg}$ is given by (\ref{grf}) with $A(x,x'; is)$ substituted 
by 
\begin{equation}
 A^{(N)}_{reg}(x,x'; is) = A(x,x'; is) - 
\sum_{k=0}^{\lfloor\frac{N}{2}\rfloor-1} a_{k}(x,x') (is)^k,
\label{regsum}
\end{equation}
one has
\begin{equation}
 \langle \upphi^{2} \rangle_{reg} = \frac{1}{(4 \pi)^{N/2}} \sum_{k=\lfloor N/2 
\rfloor}^{n'} \frac{a_{k}}{(m^{2})^{k+1-N/2}}  
\Gamma\big(k+1-\frac{N}{2}\big).
\label{sq2}
 \end{equation}
 This expression coincides with the result obtained in Ref.~\cite{LemosT}. It 
should be noted, however, that the derivation presented here is simpler. 
The vacuum polarization can be calculated once the coincidence limits
of the Hadamard-DeWitt coefficients in the concrete geometry are known.
For example, the knowledge of the coefficients  $a_{2},$ $a_{3}$ and $a_{4}$ in 
the Schwarzschild-Tangherlini spacetimes gives the field fluctuation for
$4\leq N \leq 9.$


\begin{thebibliography}{43}
\expandafter\ifx\csname natexlab\endcsname\relax\def\natexlab#1{#1}\fi
\expandafter\ifx\csname bibnamefont\endcsname\relax
  \def\bibnamefont#1{#1}\fi
\expandafter\ifx\csname bibfnamefont\endcsname\relax
  \def\bibfnamefont#1{#1}\fi
\expandafter\ifx\csname citenamefont\endcsname\relax
  \def\citenamefont#1{#1}\fi
\expandafter\ifx\csname url\endcsname\relax
  \def\url#1{\texttt{#1}}\fi
\expandafter\ifx\csname urlprefix\endcsname\relax\def\urlprefix{URL }\fi
\providecommand{\bibinfo}[2]{#2}
\providecommand{\eprint}[2][]{\url{#2}}

\bibitem[{\citenamefont{Tangherlini}(1963)}]{Tangherlini}
\bibinfo{author}{\bibfnamefont{F.}~\bibnamefont{Tangherlini}},
  \bibinfo{journal}{Nuovo Cim.} \textbf{\bibinfo{volume}{27}},
  \bibinfo{pages}{636} (\bibinfo{year}{1963}).

\bibitem[{\citenamefont{Ishibashi and Kodama}(2003)}]{Kodama}
\bibinfo{author}{\bibfnamefont{A.}~\bibnamefont{Ishibashi}} \bibnamefont{and}
  \bibinfo{author}{\bibfnamefont{H.}~\bibnamefont{Kodama}},
  \bibinfo{journal}{Prog. Theor. Phys.} \textbf{\bibinfo{volume}{110}},
  \bibinfo{pages}{901} (\bibinfo{year}{2003}), \eprint{hep-th/0305185}.

\bibitem[{\citenamefont{Schwinger}(1951)}]{Julian}
\bibinfo{author}{\bibfnamefont{J.~S.} \bibnamefont{Schwinger}},
  \bibinfo{journal}{Phys. Rev.} \textbf{\bibinfo{volume}{82}},
  \bibinfo{pages}{664} (\bibinfo{year}{1951}).

\bibitem[{\citenamefont{DeWitt}(1965)}]{Bryce1}
\bibinfo{author}{\bibfnamefont{B.~S.} \bibnamefont{DeWitt}},
  \emph{\bibinfo{title}{Dynamical Theory of groups and fields}}
  (\bibinfo{publisher}{Gordon and Breach}, \bibinfo{address}{New York},
  \bibinfo{year}{1965}).

\bibitem[{\citenamefont{DeWitt}(1975)}]{dew75}
\bibinfo{author}{\bibfnamefont{B.~S.} \bibnamefont{DeWitt}},
  \bibinfo{journal}{Phys. Reps.} \textbf{\bibinfo{volume}{19}},
  \bibinfo{pages}{295} (\bibinfo{year}{1975}).

\bibitem[{\citenamefont{Barvinsky and Vilkovisky}(1985)}]{Barvinsky}
\bibinfo{author}{\bibfnamefont{A.~O.} \bibnamefont{Barvinsky}}
  \bibnamefont{and} \bibinfo{author}{\bibfnamefont{G.~A.}
  \bibnamefont{Vilkovisky}}, \bibinfo{journal}{Phys. Rept.}
  \textbf{\bibinfo{volume}{119}}, \bibinfo{pages}{1} (\bibinfo{year}{1985}).

\bibitem[{\citenamefont{Frolov and Zel'nikov}(1984)}]{FZ3}
\bibinfo{author}{\bibfnamefont{V.~P.} \bibnamefont{Frolov}} \bibnamefont{and}
  \bibinfo{author}{\bibfnamefont{A.~I.} \bibnamefont{Zel'nikov}},
  \bibinfo{journal}{Phys. Rev.} \textbf{\bibinfo{volume}{D29}},
  \bibinfo{pages}{1057} (\bibinfo{year}{1984}).

\bibitem[{\citenamefont{Avramidi}(1989)}]{Avramidi}
\bibinfo{author}{\bibfnamefont{I.}~\bibnamefont{Avramidi}},
  \bibinfo{journal}{Theor. Math. Phys.} \textbf{\bibinfo{volume}{79}},
  \bibinfo{pages}{494} (\bibinfo{year}{1989}).

\bibitem[{\citenamefont{Taylor et~al.}(2000)\citenamefont{Taylor, Hiscock, and
  Anderson}}]{PaulA}
\bibinfo{author}{\bibfnamefont{B.~E.} \bibnamefont{Taylor}},
  \bibinfo{author}{\bibfnamefont{W.~A.} \bibnamefont{Hiscock}},
  \bibnamefont{and} \bibinfo{author}{\bibfnamefont{P.~R.}
  \bibnamefont{Anderson}}, \bibinfo{journal}{Phys. Rev.}
  \textbf{\bibinfo{volume}{D61}}, \bibinfo{pages}{084021}
  (\bibinfo{year}{2000}), \eprint{gr-qc/9911119}.

\bibitem[{\citenamefont{Sakai}(1971)}]{sakai}
\bibinfo{author}{\bibfnamefont{T.}~\bibnamefont{Sakai}},
  \bibinfo{journal}{T\^ohoku Math. J. } \textbf{\bibinfo{volume}{23}},
  \bibinfo{pages}{589} (\bibinfo{year}{1971}).

\bibitem[{\citenamefont{Gilkey}(1975)}]{gilkey}
\bibinfo{author}{\bibfnamefont{P.~B.} \bibnamefont{Gilkey}},
  \bibinfo{journal}{J. Differential Geometry} \textbf{\bibinfo{volume}{10}},
  \bibinfo{pages}{601} (\bibinfo{year}{1975}).

\bibitem[{\citenamefont{Amsterdamski et~al.}(1989)\citenamefont{Amsterdamski,
  Berkin, and O'Connor}}]{Amsterdamski}
\bibinfo{author}{\bibfnamefont{P.}~\bibnamefont{Amsterdamski}},
  \bibinfo{author}{\bibfnamefont{A.}~\bibnamefont{Berkin}}, \bibnamefont{and}
  \bibinfo{author}{\bibfnamefont{D.}~\bibnamefont{O'Connor}},
  \bibinfo{journal}{Class. Quant. Grav.} \textbf{\bibinfo{volume}{6}},
  \bibinfo{pages}{1981} (\bibinfo{year}{1989}).

\bibitem[{\citenamefont{van~de Ven}(1998)}]{Ven}
\bibinfo{author}{\bibfnamefont{A.~E.} \bibnamefont{van~de Ven}},
  \bibinfo{journal}{Class. Quant. Grav.} \textbf{\bibinfo{volume}{15}},
  \bibinfo{pages}{2311} (\bibinfo{year}{1998}), \eprint{hep-th/9708152}.

\bibitem[{\citenamefont{Frolov and Zel'nikov}(1982)}]{FZ1}
\bibinfo{author}{\bibfnamefont{V.~P.} \bibnamefont{Frolov}} \bibnamefont{and}
  \bibinfo{author}{\bibfnamefont{A.~I.} \bibnamefont{Zel'nikov}},
  \bibinfo{journal}{Phys. Lett. B} \textbf{\bibinfo{volume}{115}},
  \bibinfo{pages}{372} (\bibinfo{year}{1982}).

\bibitem[{\citenamefont{Frolov and Zel'nikov}(1983)}]{FZ2}
\bibinfo{author}{\bibfnamefont{V.~P.} \bibnamefont{Frolov}} \bibnamefont{and}
  \bibinfo{author}{\bibfnamefont{A.~I.} \bibnamefont{Zel'nikov}},
  \bibinfo{journal}{Phys. Lett. B} \textbf{\bibinfo{volume}{123}},
  \bibinfo{pages}{197} (\bibinfo{year}{1983}).

\bibitem[{\citenamefont{Matyjasek}(2000)}]{ja1}
\bibinfo{author}{\bibfnamefont{J.}~\bibnamefont{Matyjasek}},
  \bibinfo{journal}{Phys. Rev.} \textbf{\bibinfo{volume}{D61}},
  \bibinfo{pages}{124019} (\bibinfo{year}{2000}), \eprint{gr-qc/9912020}.

\bibitem[{\citenamefont{Matyjasek}(2001)}]{ja2}
\bibinfo{author}{\bibfnamefont{J.}~\bibnamefont{Matyjasek}},
  \bibinfo{journal}{Phys. Rev.} \textbf{\bibinfo{volume}{D63}},
  \bibinfo{pages}{084004} (\bibinfo{year}{2001}), \eprint{gr-qc/0010097}.

\bibitem[{\citenamefont{Anderson et~al.}(1995)\citenamefont{Anderson, Hiscock,
  and Samuel}}]{Samuel}
\bibinfo{author}{\bibfnamefont{P.~R.} \bibnamefont{Anderson}},
  \bibinfo{author}{\bibfnamefont{W.~A.} \bibnamefont{Hiscock}},
  \bibnamefont{and} \bibinfo{author}{\bibfnamefont{D.~A.}
  \bibnamefont{Samuel}}, \bibinfo{journal}{Phys.Rev.}
  \textbf{\bibinfo{volume}{D51}}, \bibinfo{pages}{4337} (\bibinfo{year}{1995}).

\bibitem[{\citenamefont{Decanini and Folacci}(2007)}]{Dec1}
\bibinfo{author}{\bibfnamefont{Y.}~\bibnamefont{Decanini}} \bibnamefont{and}
  \bibinfo{author}{\bibfnamefont{A.}~\bibnamefont{Folacci}},
  \bibinfo{journal}{Class. Quant. Grav.} \textbf{\bibinfo{volume}{24}},
  \bibinfo{pages}{4777} (\bibinfo{year}{2007}), \eprint{ArXiv:0706.0691}.

\bibitem[{\citenamefont{Decanini and Folacci}(2008)}]{Dec2}
\bibinfo{author}{\bibfnamefont{Y.}~\bibnamefont{Decanini}} \bibnamefont{and}
  \bibinfo{author}{\bibfnamefont{A.}~\bibnamefont{Folacci}},
  \bibinfo{journal}{Phys. Rev.} \textbf{\bibinfo{volume}{D78}},
  \bibinfo{pages}{044025} (\bibinfo{year}{2008}), \eprint{gr-qc/0512118}.

\bibitem[{\citenamefont{Hiscock et~al.}(1997)\citenamefont{Hiscock, Larson, and
  Anderson}}]{shane}
\bibinfo{author}{\bibfnamefont{W.~A.} \bibnamefont{Hiscock}},
  \bibinfo{author}{\bibfnamefont{S.~L.} \bibnamefont{Larson}},
  \bibnamefont{and} \bibinfo{author}{\bibfnamefont{P.~R.}
  \bibnamefont{Anderson}}, \bibinfo{journal}{Phys. Rev.}
  \textbf{\bibinfo{volume}{D56}}, \bibinfo{pages}{3571} (\bibinfo{year}{1997}),
  \eprint{gr-qc/9701004}.

\bibitem[{\citenamefont{Matyjasek et~al.}(2013)\citenamefont{Matyjasek,
  Sadurski, and Tryniecki}}]{jmP}
\bibinfo{author}{\bibfnamefont{J.}~\bibnamefont{Matyjasek}},
  \bibinfo{author}{\bibfnamefont{P.}~\bibnamefont{Sadurski}}, \bibnamefont{and}
  \bibinfo{author}{\bibfnamefont{D.}~\bibnamefont{Tryniecki}},
  \bibinfo{journal}{Phys. Rev.} \textbf{\bibinfo{volume}{D87}},
  \bibinfo{pages}{124025} (\bibinfo{year}{2013}), \eprint{ArXiv:1304.6347}.

\bibitem[{\citenamefont{Taylor et~al.}(1997)\citenamefont{Taylor, Hiscock, and
  Anderson}}]{Taylor}
\bibinfo{author}{\bibfnamefont{B.~E.} \bibnamefont{Taylor}},
  \bibinfo{author}{\bibfnamefont{W.~A.} \bibnamefont{Hiscock}},
  \bibnamefont{and} \bibinfo{author}{\bibfnamefont{P.~R.}
  \bibnamefont{Anderson}}, \bibinfo{journal}{Phys. Rev.}
  \textbf{\bibinfo{volume}{D55}}, \bibinfo{pages}{6116} (\bibinfo{year}{1997}),
  \eprint{gr-qc/9608036}.

\bibitem[{\citenamefont{Matyjasek and Sadurski}(2013)}]{frwl2013}
\bibinfo{author}{\bibfnamefont{J.}~\bibnamefont{Matyjasek}} \bibnamefont{and}
  \bibinfo{author}{\bibfnamefont{P.}~\bibnamefont{Sadurski}},
  \bibinfo{journal}{Phys. Rev.} \textbf{\bibinfo{volume}{D88}},
  \bibinfo{pages}{104015} (\bibinfo{year}{2013}), \eprint{ArXiv:1309.0552}.

\bibitem[{\citenamefont{Matyjasek et~al.}(2014)\citenamefont{Matyjasek,
  Sadurski, and Telecka}}]{frwl2014}
\bibinfo{author}{\bibfnamefont{J.}~\bibnamefont{Matyjasek}},
  \bibinfo{author}{\bibfnamefont{P.}~\bibnamefont{Sadurski}}, \bibnamefont{and}
  \bibinfo{author}{\bibfnamefont{M.}~\bibnamefont{Telecka}},
  \bibinfo{journal}{Phys. Rev.} \textbf{\bibinfo{volume}{D89}},
  \bibinfo{pages}{084055} (\bibinfo{year}{2014}).

\bibitem[{\citenamefont{Kofman and Sahni}(1983)}]{Kofman1}
\bibinfo{author}{\bibfnamefont{L.~A.} \bibnamefont{Kofman}} \bibnamefont{and}
  \bibinfo{author}{\bibfnamefont{V.}~\bibnamefont{Sahni}},
  \bibinfo{journal}{Phys. Lett.} \textbf{\bibinfo{volume}{B127}},
  \bibinfo{pages}{197} (\bibinfo{year}{1983}).

\bibitem[{\citenamefont{Sahni and Kofman}(1986)}]{Kofman2}
\bibinfo{author}{\bibfnamefont{V.}~\bibnamefont{Sahni}} \bibnamefont{and}
  \bibinfo{author}{\bibfnamefont{L.~A.} \bibnamefont{Kofman}},
  \bibinfo{journal}{Phys. Lett.} \textbf{\bibinfo{volume}{A117}},
  \bibinfo{pages}{275} (\bibinfo{year}{1986}).

\bibitem[{\citenamefont{Matyjasek and Tryniecki}(2009{\natexlab{a}})}]{Mata42}
\bibinfo{author}{\bibfnamefont{J.}~\bibnamefont{Matyjasek}} \bibnamefont{and}
  \bibinfo{author}{\bibfnamefont{D.}~\bibnamefont{Tryniecki}},
  \bibinfo{journal}{Mod. Phys. Lett.} \textbf{\bibinfo{volume}{A24}},
  \bibinfo{pages}{2517} (\bibinfo{year}{2009}{\natexlab{a}}),
  \eprint{ArXiv:0908.2648}.

\bibitem[{\citenamefont{Thompson and Lemos}(2009)}]{LemosT}
\bibinfo{author}{\bibfnamefont{R.}~\bibnamefont{Thompson}} \bibnamefont{and}
  \bibinfo{author}{\bibfnamefont{J.~P.} \bibnamefont{Lemos}},
  \bibinfo{journal}{Phys. Rev.} \textbf{\bibinfo{volume}{D80}},
  \bibinfo{pages}{064017} (\bibinfo{year}{2009}), \eprint{ArXiv:0811.3962}.

\bibitem[{\citenamefont{Kent and Winstanley}(2014)}]{Kent}
\bibinfo{author}{\bibfnamefont{C.}~\bibnamefont{Kent}} \bibnamefont{and}
  \bibinfo{author}{\bibfnamefont{E.}~\bibnamefont{Winstanley}}
  (\bibinfo{year}{2014}), \eprint{ArXiv:1408.6738}.

\bibitem[{\citenamefont{Vermaseren}(2000)}]{Vermaseren}
\bibinfo{author}{\bibfnamefont{J.}~\bibnamefont{Vermaseren}}
  (\bibinfo{year}{2000}), \eprint{math-ph/0010025}.

\bibitem[{\citenamefont{Tentyukov and Vermaseren}(2010)}]{Tentyukov}
\bibinfo{author}{\bibfnamefont{M.}~\bibnamefont{Tentyukov}} \bibnamefont{and}
  \bibinfo{author}{\bibfnamefont{J.}~\bibnamefont{Vermaseren}},
  \bibinfo{journal}{Comput. Phys. Commun.} \textbf{\bibinfo{volume}{181}},
  \bibinfo{pages}{1419} (\bibinfo{year}{2010}), \eprint{hep-ph/0702279}.

\bibitem[{\citenamefont{Kuipers et~al.}(2013)\citenamefont{Kuipers, Ueda,
  Vermaseren, and Vollinga}}]{Kuipers}
\bibinfo{author}{\bibfnamefont{J.}~\bibnamefont{Kuipers}},
  \bibinfo{author}{\bibfnamefont{T.}~\bibnamefont{Ueda}},
  \bibinfo{author}{\bibfnamefont{J.}~\bibnamefont{Vermaseren}},
  \bibnamefont{and} \bibinfo{author}{\bibfnamefont{J.}~\bibnamefont{Vollinga}},
  \bibinfo{journal}{Comput. Phys. Commun.} \textbf{\bibinfo{volume}{184}},
  \bibinfo{pages}{1453} (\bibinfo{year}{2013}), \eprint{ArXiv:1203.6543}.

\bibitem[{\citenamefont{York}(1985)}]{York}
\bibinfo{author}{\bibfnamefont{J.~W.} \bibnamefont{York}},
  \bibinfo{journal}{Phys. Rev.} \textbf{\bibinfo{volume}{D31}},
  \bibinfo{pages}{775} (\bibinfo{year}{1985}).

\bibitem[{\citenamefont{Lousto and Sanchez}(1988)}]{Sanchez}
\bibinfo{author}{\bibfnamefont{C.}~\bibnamefont{Lousto}} \bibnamefont{and}
  \bibinfo{author}{\bibfnamefont{N.~G.} \bibnamefont{Sanchez}},
  \bibinfo{journal}{Phys. Lett.} \textbf{\bibinfo{volume}{B212}},
  \bibinfo{pages}{411} (\bibinfo{year}{1988}).

\bibitem[{\citenamefont{Hochberg et~al.}(1993)\citenamefont{Hochberg, Kephart,
  and York}}]{Hochberg1}
\bibinfo{author}{\bibfnamefont{D.}~\bibnamefont{Hochberg}},
  \bibinfo{author}{\bibfnamefont{T.~W.} \bibnamefont{Kephart}},
  \bibnamefont{and} \bibinfo{author}{\bibfnamefont{J.}~\bibnamefont{York},
  \bibfnamefont{James~W.}}, \bibinfo{journal}{Phys. Rev.}
  \textbf{\bibinfo{volume}{D48}}, \bibinfo{pages}{479} (\bibinfo{year}{1993}),
  \eprint{gr-qc/9211009}.

\bibitem[{\citenamefont{Hochberg and Kephart}(1993)}]{Hochberg2}
\bibinfo{author}{\bibfnamefont{D.}~\bibnamefont{Hochberg}} \bibnamefont{and}
  \bibinfo{author}{\bibfnamefont{T.~W.} \bibnamefont{Kephart}},
  \bibinfo{journal}{Phys. Rev.} \textbf{\bibinfo{volume}{D47}},
  \bibinfo{pages}{1465} (\bibinfo{year}{1993}), \eprint{gr-qc/9211008}.

\bibitem[{\citenamefont{Fursaev}(1995)}]{Fursaev}
\bibinfo{author}{\bibfnamefont{D.~V.} \bibnamefont{Fursaev}},
  \bibinfo{journal}{Phys. Rev.} \textbf{\bibinfo{volume}{D51}},
  \bibinfo{pages}{5352} (\bibinfo{year}{1995}), \eprint{hep-th/9412161}.

\bibitem[{\citenamefont{Banerjee and Majhi}(2008)}]{Banerjee}
\bibinfo{author}{\bibfnamefont{R.}~\bibnamefont{Banerjee}} \bibnamefont{and}
  \bibinfo{author}{\bibfnamefont{B.~R.} \bibnamefont{Majhi}},
  \bibinfo{journal}{Phys. Lett.} \textbf{\bibinfo{volume}{B662}},
  \bibinfo{pages}{62} (\bibinfo{year}{2008}), \eprint{ArXiv:0801.0200}.

\bibitem[{\citenamefont{Banerjee et~al.}(2010)\citenamefont{Banerjee, Kiefer,
  and Majhi}}]{Banerjee2}
\bibinfo{author}{\bibfnamefont{R.}~\bibnamefont{Banerjee}},
  \bibinfo{author}{\bibfnamefont{C.}~\bibnamefont{Kiefer}}, \bibnamefont{and}
  \bibinfo{author}{\bibfnamefont{B.~R.} \bibnamefont{Majhi}},
  \bibinfo{journal}{Phys. Rev.} \textbf{\bibinfo{volume}{D82}},
  \bibinfo{pages}{044013} (\bibinfo{year}{2010}), \eprint{ArXiv:1005.2264}.

\bibitem[{\citenamefont{Matyjasek and
  Tryniecki}(2009{\natexlab{b}})}]{Tryniecki}
\bibinfo{author}{\bibfnamefont{J.}~\bibnamefont{Matyjasek}} \bibnamefont{and}
  \bibinfo{author}{\bibfnamefont{D.}~\bibnamefont{Tryniecki}},
  \bibinfo{journal}{Phys. Rev.} \textbf{\bibinfo{volume}{D79}},
  \bibinfo{pages}{084017} (\bibinfo{year}{2009}{\natexlab{b}}),
  \eprint{ArXiv:0901.2746}.

\bibitem[{\citenamefont{Matyjasek and Zwierzchowska}(2012)}]{Kasia}
\bibinfo{author}{\bibfnamefont{J.}~\bibnamefont{Matyjasek}} \bibnamefont{and}
  \bibinfo{author}{\bibfnamefont{K.}~\bibnamefont{Zwierzchowska}},
  \bibinfo{journal}{Phys. Rev.} \textbf{\bibinfo{volume}{D85}},
  \bibinfo{pages}{024009} (\bibinfo{year}{2012}), \eprint{ArXiv:1110.0041}.

\bibitem[{\citenamefont{Myers and Perry}(1986)}]{Myers}
\bibinfo{author}{\bibfnamefont{R.~C.} \bibnamefont{Myers}} \bibnamefont{and}
  \bibinfo{author}{\bibfnamefont{M.}~\bibnamefont{Perry}},
  \bibinfo{journal}{Annals Phys.} \textbf{\bibinfo{volume}{172}},
  \bibinfo{pages}{304} (\bibinfo{year}{1986}).

\end{thebibliography}
\end{document}